\documentclass[3p,number,sort&compress]{elsarticle}
\usepackage{amsmath,amssymb,bm,graphicx}
\newcommand{\mi}{\mathrm{i}}
\newcommand{\gagg}{g_{a\gamma\gamma}}

\journal{Journal of Computational Physics}

\begin{document}

\begin{frontmatter}
\title{$X$-dispersionless solver for electromagnetic and axion fields in a cylindrical particle-in-cell code}

\author[addr1,addr2]{Xiangyan An}
\author[addr2]{Min Chen\corref{cor1}}
\ead{minchen@sjtu.edu.cn}
\author[addr1]{Yipeng Wu}
\author[addr1]{Jianglai Liu}

\author[addr1,addr2]{Zhengming Sheng}
\author[addr1,addr2]{Jie Zhang}

\cortext[cor1]{Corresponding author}

\address[addr1]{State Key Laboratory of Dark Matter Physics, Tsung-Dao Lee Institute, Shanghai Jiao Tong University, Shanghai 201210, China}
\address[addr2]{State Key Laboratory of Dark Matter Physics, Key Laboratory for Laser Plasmas (MoE), School of Physics and Astronomy, Shanghai Jiao Tong University, Shanghai 200240, China}

\begin{abstract}
We develop a quasi-cylindrical direction-splitting (QDS) dispersionless Maxwell solver for the open-source EPOCH particle-in-cell framework. The method preserves the exact axial phase velocity for every retained azimuthal mode by advecting transverse transport variables with the $c\Delta t=\Delta x$ shift. Building on the same transport viewpoint, we formulate an axion solver that advances the Klein--Gordon equation and the axion-regenerated electromagnetic fields within the QDS solver, so that axion--photon coupling is handled self-consistently. Benchmarks demonstrate that the cylindrical QDS solver reproduces the standard Yee wakefield structure while removing the axial group-velocity error and reducing computational cost by orders of magnitude. In the resonant axion generation test via two-color laser mixing, the QDS solver eliminates the spurious vacuum source term and maintains phase matching, whereas the Yee solver suffers from rapid dephasing due to numerical dispersion. The proposed algorithm therefore enables quasi-3D simulations of axion--photon physics in laser-plasma interaction with dispersion-free laser transport.
\end{abstract}

\end{frontmatter}

\section{Introduction}
Particle-in-cell (PIC) simulations are widely used for modeling plasma dynamics, laser--plasma interaction, and plasma-based particle acceleration across scales that are otherwise inaccessible to analytic theory or fluid models~\cite{Lifschitz2009,Fonseca2008,Pukhov2016,Arber2015,Vay2013,Derouillat2018,Burau2010}. In explicit PIC algorithms, Maxwell's equations are most often advanced on the second-order Yee finite-difference time-domain grid~\cite{Yee1966}, which provides a second-order accurate explicit update on a compact stencil. Nevertheless, the Yee lattice suffers from numerical dispersion. Electromagnetic (EM) waves propagate more slowly than $c$ on the mesh so that relativistic beams or boosted-frame simulations can excite spurious numerical Cherenkov instability (NCI) modes~\cite{Godfrey2013,Vay2011,Pukhov2020}. In laser plasma wakefield acceleration, the resulting phase errors degrade laser group velocities, distort wake structures, and generate nonphysical phenomena.

Several strategies have been advanced to reduce the numerical dispersion and NCI, including spectral solvers~\cite{lehe_spectral_2016}, Galilean-frame formulations~\cite{Lehe2016}, hybrid stencils that increase the effective Courant number~\cite{Godfrey2013}, and direction-splitting Maxwell updates in PIC codes~\cite{Nuter2014}. Among them, the $X$-dispersionless solver proposed by Pukhov~\cite{Pukhov2020} transports combinations of transverse electromagnetic fields by one cell per step along the principal axis, thereby eliminating the axial phase error while retaining a compact real-space stencil. Pukhov (2020) termed this framework the rhombi-in-plane (RIP) solver. In the present quasi-cylindrical formulation we retain its axial dispersionless transport idea and reconstruct the transverse discretization under cylindrical mode decomposition and staggered grids. This approach enables high-accuracy simulation of electromagnetic-wave propagation at comparatively low resolutions, thereby reducing computational requirements.

On the other hand, quasi-cylindrical PIC codes, introduced by Lifschitz {et al.}~\cite{Lifschitz2009}, expand fields and currents in azimuthal modes to capture essential three-dimensional physics at near-2D computational expense. The EPOCH code implements this concept, incorporating relativistic particle pusher, charge-conserving current deposition, and a suite of practical diagnostic tools~\cite{Arber2015,cylindrical_epoch_github}. However, its cylindrical Maxwell solver still inherits the Yee dispersion along the propagation direction, thereby retaining the same limitations outlined above. A dispersionless update that remains compatible with the Lifschitz staggering would be valuable for high-accuracy simulations of long-distance laser propagation and plasma-based accelerator stages.

Moreover, the axion was proposed to resolve the strong CP problem~\cite{Peccei1977a,Peccei1977b,Wilczek1978,Weinberg1978}, and both axions and axion-like particles have emerged as possible dark-matter candidates~\cite{Preskill1983,Abbott1983,Dine1983}. This dual role has boosted laboratory searches including light shining through a wall (LSW)~\cite{Sikivie1983,van_bibber_proposed_1987} and some recent concepts based on the laser--plasma interactions~\cite{Tercas2018,Burton2018,Mendonca2020a,Mendonca2020b,Huang2022,an_situ_2025,an_enhanced_2025}. Because the axion--photon coupling is extremely weak, efficient conversion between photons and axions through laser-plasma interactions requires a long interaction distance, while maintaining the relative phase between the laser and axion fields within one wavelength in the simulation. These constraints make the combination of quasi-cylindrical geometry and a dispersionless quasi-cylindrical direction-splitting (QDS) solver a promising approach for conducting realistic design studies.

In this paper, we extend Pukhov's RIP concept to the cylindrical-coordinate branch of EPOCH. Each retained azimuthal mode is updated with transport variables that move exactly at $c$ along the axial direction. We also derive a matched axion transport update and an axion-regenerated electromagnetic (AREM) update, ensuring the coupled Klein--Gordon and Maxwell equations remain compatible with the QDS solver. The resulting solver reproduces the standard Yee results for the laser wakefield shape and amplitude, yet it maintains the correct laser group velocity at a low axial resolution and provides accurate results of axion conversion.

\section{Numerical Method}\label{sec:method}
In this section, we formulate the QDS update schemes for the electromagnetic fields and the axion field in cylindrical coordinates. The cylindrical geometry is handled via a truncated Fourier decomposition in the azimuthal direction ($\theta$):
\begin{equation}\label{eq:fourier_decomp}
F(x,r,\theta,t)=\text{Re}\left\{\sum_{m=0}^{m_{\max}} F^{(m)}(x,r,t)e^{-\mi m \theta}\right\}=\frac{1}{2}\sum_{m=0}^{m_{\max}}\left[F^{(m)}e^{-\mi m\theta}+F^{(-m)}e^{\mi m\theta}\right],
\end{equation}
where $F^{(-m)}=[F^{(m)}]^*$ enforces real-valued fields and $m_{\max}$ is the highest retained mode. Maxwell's equations undergo decoupling into $(m_{\max}+1)$-modes under this representation. Below we focus on a single mode $m$, dropping the superscript $m$ when it is unambiguous.

\subsection{Electromagnetic field QDS solver in cylindrical coordinates}
The mode-decomposed Maxwell's equations in cylindrical coordinates $(x,r)$ for mode $m$ are (in SI units):
\begin{align}
\partial_t B_x^{(m)}&=-\frac{1}{r}\partial_r\!\big(r E_\theta^{(m)}\big)-\frac{\mi m}{r}E_r^{(m)}, \label{eq:maxwell1}\\
\partial_t B_r^{(m)}&=\partial_x E_\theta^{(m)}+\frac{\mi m}{r}E_x^{(m)}, \\
\partial_t B_\theta^{(m)}&=\partial_r E_x^{(m)}-\partial_x E_r^{(m)}, \\
\partial_t E_x^{(m)}&=\frac{c^2}{r}\partial_r\!\big(r B_\theta^{(m)}\big)+\frac{\mi m c^2}{r}B_r^{(m)}-\frac{1}{\varepsilon_0}j_x^{(m)}, \\
\partial_t E_r^{(m)}&=-c^2\partial_x B_\theta^{(m)}-\frac{\mi m c^2}{r}B_x^{(m)}-\frac{1}{\varepsilon_0}j_r^{(m)}, \\
\partial_t E_\theta^{(m)}&=c^2\partial_x B_r^{(m)}-c^2\partial_r B_x^{(m)}-\frac{1}{\varepsilon_0}j_\theta^{(m)}, \label{eq:maxwell6}
\end{align}
where $\mathbf{j}^{(m)}=(j_x^{(m)},j_r^{(m)},j_\theta^{(m)})$ is the Fourier component of the current density. The goal of the QDS scheme is to remove the numerical dispersion along $x$ direction, treating the $x$-propagation of electromagnetic waves implicitly. To achieve this, we combine certain $E$ and $B$ components into transport variables that represent waves propagating forward or backward along the $x$ direction. Following the strategy of the planar QDS solver, we define:
\begin{equation}\label{eq:Tdefs}
T_r^\pm=E_r^{(m)}\pm c B_\theta^{(m)}, \qquad
T_\theta^\pm=E_\theta^{(m)}\pm c B_r^{(m)}~.
\end{equation}
Using Eqs.~\eqref{eq:maxwell1}--\eqref{eq:maxwell6}, one can derive evolution equations for these transport variables. Substituting the time derivatives of $E_r^{(m)}$ and $B_\theta^{(m)}$ into $\partial_t T_r^\pm$ and similarly for $T_\theta^\pm$, one can obtain:
\begin{align}
\partial_t T_r^\pm \pm c\partial_x T_r^\pm&=-\frac{\mi m c^2}{r}B_x-\frac{1}{\varepsilon_0}j_r\pm c\partial_r E_x, \label{eq:Tr_propagation} \\
\partial_t T_\theta^\pm \mp c\partial_x T_\theta^\pm&=-c^2\partial_r B_x-\frac{1}{\varepsilon_0}j_\theta\pm \frac{\mi m c}{r}E_x, \label{eq:Ttheta_propagation}
\end{align}
while the $x$-components $E_x$ and $B_x$ evolve according to:
\begin{equation}\label{eq:longitudinal_update}
\partial_t E_x=\frac{c^2}{r}\partial_r\!\big(r B_\theta\big)+\frac{\mi m c^2}{r}B_r-\frac{1}{\varepsilon_0}j_x\equiv \Gamma_x,
\qquad
\partial_t B_x=-\frac{1}{r}\partial_r\!\big(r E_\theta\big)-\frac{\mi m}{r}E_r\equiv \Phi_x,
\end{equation}
where we have introduced shorthand notations $\Gamma_x$ and $\Phi_x$ to denote the right-hand sides. For convenience, we also define:
\begin{equation}
\Gamma_r=-\frac{\mi m c^2}{r}B_x-\frac{1}{\varepsilon_0}j_r, \qquad
\Gamma_\theta=-c^2\partial_r B_x-\frac{1}{\varepsilon_0}j_\theta, \qquad
\Phi_r=\partial_r E_x, \qquad
\Phi_\theta=\frac{\mi m}{r}E_x,
\end{equation}
to simplify notation in Eqs.~\eqref{eq:Tr_propagation}--\eqref{eq:Ttheta_propagation}. In terms of these, the transport equations can be concisely written as:
\begin{equation}\label{eq:transport_compact}
\partial_t T_r^\pm \pm c\partial_x T_r^\pm = \Gamma_r \pm c\Phi_r, \qquad
\partial_t T_\theta^\pm \mp c\partial_x T_\theta^\pm = \Gamma_\theta \pm c\Phi_\theta~,
\end{equation}
with $\partial_t E_x = \Gamma_x$ and $\partial_t B_x = \Phi_x$ as given above.

\begin{figure}[!tb]
\centering
\includegraphics[width=0.8\textwidth]{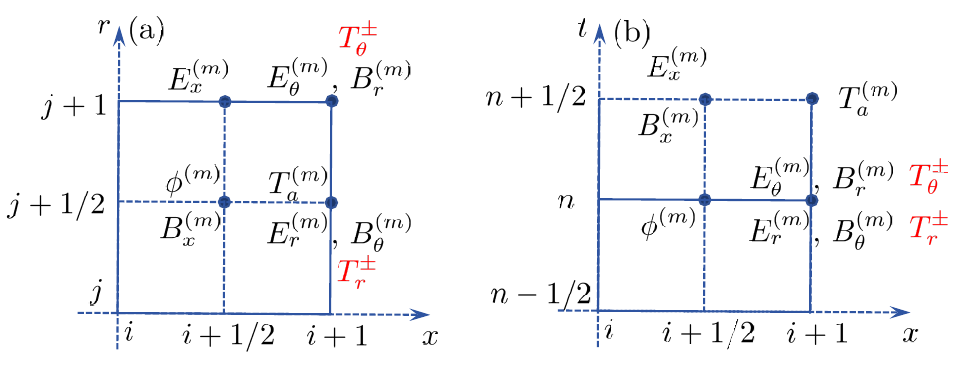}
\caption{Staggered grid configuration of field components in cylindrical coordinates. (a) Field configuration in $x$-$r$ plane. (b) Field configuration in $x$-$t$ plane.}
\label{fig:stagger}
\end{figure}

Equations \eqref{eq:transport_compact} represent a pair of advection equations for $T_r^\pm$ and $T_\theta^\pm$ along the $x$-axis (with speeds $\pm c$), supplemented by source terms $\Gamma_{r,\theta}$ and $\Phi_{r,\theta}$. The key to the QDS scheme is to solve these equations in an implicit (or effectively semi-implicit) manner such that a wave traveling along $+x$ or $-x$ experiences zero numerical phase error. Assuming a grid spacing $\Delta x$ and time step $\Delta t$ satisfying the special condition $c\Delta t = \Delta x$, the update for the $T$-fields over a full step reads:
\begin{align}
T_{r,\,i+1,\,j+1/2}^{+,\,n+1}&=T_{r,\,i,\,j+1/2}^{+,\,n}+\Delta t\big[\Gamma_r+c\Phi_r\big]_{\,i+1/2,\,j+1/2}^{\,n+1/2}, \label{eq:Trp_update}\\
T_{r,\,i+1,\,j+1/2}^{-,\,n+1}&=T_{r,\,i+2,\,j+1/2}^{-,\,n}+\Delta t\big[\Gamma_r-c\Phi_r\big]_{\,i+3/2,\,j+1/2}^{\,n+1/2}, \label{eq:Trm_update}\\
T_{\theta,\,i+1,\,j+1}^{+,\,n+1}&=T_{\theta,\,i+2,\,j+1}^{+,\,n}+\Delta t\big[\Gamma_\theta+c\Phi_\theta\big]_{\,i+3/2,\,j+1}^{\,n+1/2}, \label{eq:Ttp_update}\\
T_{\theta,\,i+1,\,j+1}^{-,\,n+1}&=T_{\theta,\,i,\,j+1}^{-,\,n}+\Delta t\big[\Gamma_\theta-c\Phi_\theta\big]_{\,i+1/2,\,j+1}^{\,n+1/2}, \label{eq:Ttm_update}
\end{align}
where $i$ indexes cell centers along the $x$-axis, $j$ denotes the radial positions, and $n$ labels the integer time levels of the transverse fields. The source terms $\Gamma$ and $\Phi$ are evaluated at the intermediate positions indicated by the subscripts (e.g. the $i+1/2$ corresponds to the $x$-faces at which the fields $E_x$ and $B_x$ are collocated) and at the intermediate time $t^{n+1/2}$.

In accordance with the staggering configuration shown in Fig.~\ref{fig:stagger}, these source terms take the explicit forms
\begin{align}
\Gamma_{r,\,i+1/2,\,j+1/2}^{\,n+1/2} &=-\frac{\mi m c^2}{r_{j+1/2}} B_{x,\,i+1/2,\,j+1/2}^{\,n+1/2}-\frac{1}{\varepsilon_0}j_{r,\,i+1/2,\,j+1/2}^{\,n+1/2},\\
\Phi_{r,\,i+1/2,\,j+1/2}^{\,n+1/2} &= \big(\partial_r E_x\big)_{\,i+1/2,\,j+1/2}^{\,n+1/2},\\
\Gamma_{\theta,\,i+1/2,\,j}^{\,n+1/2} &=-c^2\big(\partial_r B_x\big)_{\,i+1/2,\,j}^{\,n+1/2}-\frac{1}{\varepsilon_0}j_{\theta,\,i+1/2,\,j}^{\,n+1/2},\\
\Phi_{\theta,\,i+1/2,\,j}^{\,n+1/2} &= \frac{\mi m}{r_j}E_{x,\,i+1/2,\,j}^{\,n+1/2},
\end{align}
where $r_{j}$ and $r_{j+1/2}$ denote the corresponding radial coordinates. In the above equations the radial derivatives are evaluated with centered differences along $r$.

After updating the transport variables $T_r^\pm$ and $T_\theta^\pm$ to the new time $t^{n+1}$, the electromagnetic field components can be reconstructed at their native locations. From Eq.~\eqref{eq:Tdefs}, one can obtain:
\begin{equation}\label{eq:reconstruct_fields}
E_{r,\,i+1,\,j+1/2}^{\,n} = \frac{T_{r,\,i+1,\,j+1/2}^{+,\,n} + T_{r,\,i+1,\,j+1/2}^{-,\,n}}{2}, \quad
B_{\theta,\,i+1,\,j+1/2}^{\,n} = \frac{T_{r,\,i+1,\,j+1/2}^{+,\,n} - T_{r,\,i+1,\,j+1/2}^{-,\,n}}{2c},
\end{equation}
\begin{equation*}
E_{\theta,\,i+1,\,j+1}^{\,n} = \frac{T_{\theta,\,i+1,\,j+1}^{+,\,n} + T_{\theta,\,i+1,\,j+1}^{-,\,n}}{2}, \quad
B_{r,\,i+1,\,j+1}^{\,n} = \frac{T_{\theta,\,i+1,\,j+1}^{+,\,n} - T_{\theta,\,i+1,\,j+1}^{-,\,n}}{2c}~.
\end{equation*}
Finally, the longitudinal fields $E_x$ and $B_x$ are updated using Eq.~\eqref{eq:longitudinal_update} with a central difference in time at their staggered locations:
\begin{equation}\label{eq:Ex_update}
E_{x,\,i+1/2,\,j+1}^{\,n+1/2} = E_{x,\,i+1/2,\,j+1}^{\,n-1/2} + \Delta t\,\Gamma_{x,\,i+1/2,\,j+1}^{\,n}, \qquad
B_{x,\,i+1/2,\,j+1/2}^{\,n+1/2} = B_{x,\,i+1/2,\,j+1/2}^{\,n-1/2} + \Delta t\,\Phi_{x,\,i+1/2,\,j+1/2}^{\,n},
\end{equation}
where $\Gamma_{x}^{n}$ and $\Phi_{x}^{n}$ are constructed from the transverse fields at time $t^{n}$ (readily obtained from the $T$-variables on the integer grid). This completes one full update cycle for the electromagnetic fields.

Referring again to Fig.~\ref{fig:stagger}, the collocated expressions are
\begin{align}
\Gamma_{x,\,i+1/2,\,j}^{\,n} &= \frac{c^2}{r_j}\,\partial_r\!\big[r B_{\theta}\big]_{\,i+1/2,\,j}^{\,n}+\frac{\mi m c^2}{r_j}B_{r,\,i+1/2,\,j}^{\,n}-\frac{1}{\varepsilon_0}j_{x,\,i+1/2,\,j}^{\,n}, \label{eq:Gammax_explicit} \\
\Phi_{x,\,i+1/2,\,j+1/2}^{\,n} &=-\frac{1}{r_{j+1/2}}\,\partial_r\!\big[r E_{\theta}\big]_{\,i+1/2,\,j+1/2}^{\,n}-\frac{\mi m}{r_{j+1/2}}E_{r,\,i+1/2,\,j+1/2}^{\,n}, \label{eq:Phix_explicit}
\end{align}
where the radial derivatives are evaluated via centered differences aligned with the local staggered grid positions. As an example, we consider the evaluation of $r^{-1}\partial_r(rB_\theta)$ at $\left(i+\tfrac{1}{2},j\right)$ even though $B_\theta$ is naturally stored at $\left(i+1,j+\tfrac{1}{2}\right)$. We first rewrite
\begin{equation*}
\frac{1}{r}\partial_r\big(rB_\theta\big)=\frac{B_\theta}{r}+\partial_r B_\theta
\end{equation*}
and obtain the two pieces separately at the required location. The value of $B_\theta$ is averaged onto $\left(i+\tfrac{1}{2},j\right)$ with a four-point stencil,
\begin{equation*}
B_{\theta,\,i+1/2,\,j}=\frac{1}{4}\left[B_{\theta,\,i,\,j-1/2}+B_{\theta,\,i,\,j+1/2}+B_{\theta,\,i+1,\,j-1/2}+B_{\theta,\,i+1,\,j+1/2}\right],
\end{equation*}
while the radial derivative is evaluated using a centered difference constructed from the same neighboring samples,
\begin{equation*}
\big(\partial_r B_\theta\big)_{\,i+1/2,\,j}=\frac{\big[B_{\theta,\,i,\,j+1/2}+B_{\theta,\,i+1,\,j+1/2}\big]-\big[B_{\theta,\,i,\,j-1/2}+B_{\theta,\,i+1,\,j-1/2}\big]}{2\Delta r}~.
\end{equation*}
Substituting these two expressions yields the desired term $r^{-1}\partial_r(rB_\theta)$ at the position $\left(i+\tfrac{1}{2},j\right)$. For all other cases where a field must be sampled away from its native staggered grid location, the same methodology is applied. Centered averages provide the field value, and centered differences provide the required derivatives.

\subsection{Axion field QDS solver}
We now formulate the solver for the axion field $\phi(x,r,t)$, which satisfies a wave equation with a source from the electromagnetic fields~\cite{van_bibber_proposed_1987,Tercas2018,an_modeling_2024}. The axion field equation is:
\begin{equation}\label{eq:axion_waveeq}
\left(\frac{1}{c^2}\frac{\partial^2}{\partial t^2} - \nabla^2 + \frac{m_a^2 c^2}{\hbar^2}\right)\phi = \frac{\gagg}{\hbar \mu_0}\,\mathbf{E}\cdot\mathbf{B}~,
\end{equation}
where $m_a$ is the axion mass and $\gagg$ is the coupling constant. Because the cylindrical variant of EPOCH stores azimuthal modes rather than the full $\theta$-resolved fields, the source term $S_a=\mathbf{E}\cdot\mathbf{B}$ must be projected mode-by-mode using Eq.~\eqref{eq:fourier_decomp} before it is injected into Eq.~\eqref{eq:axion_waveeq}. Applying the expansion to $S_a=\sum_{\alpha}E_\alpha B_\alpha$ (with $\alpha\in\{x,r,\theta\}$) yields a set of coupled convolution sums. For $m=0$ the real-valued source reads
\begin{equation}\label{eq:source_mode_zero}
S_a^{(0)}=\sum_{\alpha}\left[E_\alpha^{(0)}B_\alpha^{(0)}+\frac{1}{2}\sum_{k>0}\text{Re}\big\{E_\alpha^{(k)}B_\alpha^{(-k)}\big\}\right],
\end{equation}
whereas for $m>0$ we obtain
\begin{equation}\label{eq:source_mode_m}
S_a^{(m)}=\frac{1}{2}\sum_{\alpha}\Bigg[\sum_{k+l=m}E_\alpha^{(k)}B_\alpha^{(l)}+\sum_{k-l=m}E_\alpha^{(k)}B_\alpha^{(-l)}+\sum_{l-k=m}E_\alpha^{(-k)}B_\alpha^{(l)}\Bigg],
\end{equation}
with all indices $k,l\ge 0$. For compactness we denote the bracketed sums as
\begin{equation}\label{eq:mode_convolution}
\mathcal{C}_m\{A,B\} \equiv
\begin{cases}
\displaystyle A^{(0)}B^{(0)}+\frac{1}{2}\sum_{k>0}\text{Re}\big\{A^{(k)}B^{(-k)}\big\}, & m=0,\\
\displaystyle \frac{1}{2}\Bigg[\sum_{k+l=m}A^{(k)}B^{(l)}+\sum_{k-l=m}A^{(k)}B^{(-l)}+\sum_{l-k=m}A^{(-k)}B^{(l)}\Bigg], & m>0,
\end{cases}
\end{equation}
so that $S_a^{(m)}=\sum_{\alpha}\mathcal{C}_m\{E_\alpha,B_\alpha\}$. Equations \eqref{eq:source_mode_zero}--\eqref{eq:source_mode_m} make explicit that only $(k,\,l)$ mode pairs satisfying $m=\pm k\pm l$ contribute to the axion source for mode $m$.
We treat the resulting Klein-Gordon-type wave equation (\ref{eq:axion_waveeq}) with a similar strategy as the electromagnetic fields by splitting $\phi$ into components that propagate along $+x$ and $-x$ directions. Specifically, we define two axion transport variables:
\begin{equation}\label{eq:T_axion_def}
T_a^+ = \frac{1}{c}\partial_t \phi - \partial_x \phi~, \qquad
T_a^- = \frac{1}{c}\partial_t \phi + \partial_x \phi~.
\end{equation}
These can be viewed as analogs of right-propagating and left-propagating wave components of $\phi$. In terms of $T_a^\pm$, one can show that the wave equation \eqref{eq:axion_waveeq} is equivalent to:
\begin{equation}\label{eq:axion_transport_eqs}
\frac{1}{c}\partial_t T_a^+ + \partial_x T_a^+ = \Gamma_a^{(m)}~, \qquad
\frac{1}{c}\partial_t T_a^- - \partial_x T_a^- = \Gamma_a^{(m)}~,
\end{equation}
where
$\Gamma_a^{(m)}(x,r,t) \equiv \left(\partial_r^2 + \frac{1}{r}\partial_r - \frac{m^2}{r^2}\right)\phi^{(m)} - \frac{m_a^2 c^2}{\hbar^2}\phi^{(m)} + \frac{\gagg}{\hbar \mu_0}S_a^{(m)}$
represents the non-advective terms (i.e. transverse Laplacian rendered explicitly in cylindrical coordinates for mode $m$, mass term, and source coupling to EM fields). The form of Eq.~\eqref{eq:axion_transport_eqs} makes it clear that $T_a^+$ and $T_a^-$ propagate along $+x$ and $-x$ respectively, with speed $c$, while sourcing each other through $\Gamma_a^{(m)}$.

We discretize the axion transport equations \eqref{eq:axion_transport_eqs} using an approach analogous to that employed for the EM solver. The scalar field $\phi$ is stored at $(i+1/2,\,j+1/2,\,n)$. The transport variables $T_a^\pm$ occupy the integer $x$ position $i$, share the same radial offset $j+1/2$, and live at half-integer times $n+1/2$. Using central differencing (again with $c\Delta t = \Delta x$), we update the axion transport variables as:
\begin{align}
T_{a,\,i+1,\,j+1/2}^{+,\,n+1/2} &= T_{a,\,i,\,j+1/2}^{+,\,n-1/2} + c\Delta t\,\Gamma_{a,\,i+1/2,\,j+1/2}^{(m),\,n}, \label{eq:Ta_plus_update}\\
T_{a,\,i,\,j+1/2}^{-,\,n+1/2}   &= T_{a,\,i+1,\,j+1/2}^{-,\,n-1/2} + c\Delta t\,\Gamma_{a,\,i+1/2,\,j+1/2}^{(m),\,n}, \label{eq:Ta_minus_update}
\end{align}
where $\Gamma_{a,\,i+1/2,\,j+1/2}^{(m),n}$ is the source term evaluated at the same $(i+1/2, j+1/2)$ location as $\phi^{(m)}$. Thus $T_a^+$ values shift one cell to the right per full time step while $T_a^-$ shift one cell to the left, carrying the axion information along $x$ without dispersion. The pointwise relation is
\begin{equation}
\begin{split}
\Gamma_{a,\,i+1/2,\,j+1/2}^{(m),\,n}&=\big(\partial_r^2\phi^{(m)}\big)_{\,i+1/2,\,j+1/2}^{\,n}+\frac{1}{r_{j+1/2}}\big(\partial_r \phi^{(m)}\big)_{\,i+1/2,\,j+1/2}^{\,n} \\
&\phantom{=}-\frac{m^2}{r_{j+1/2}^2}\phi_{\,i+1/2,\,j+1/2}^{(m),\,n}-\frac{m_a^2 c^2}{\hbar^2}\phi_{\,i+1/2,\,j+1/2}^{(m),\,n}+\frac{\gagg}{\hbar \mu_0}S_{a,\,i+1/2,\,j+1/2}^{(m),\,n},
\end{split}
\end{equation}
which follows directly from the placement of the axion field $\phi$ and the axion transport variables $T_a^\pm$ in Fig.~\ref{fig:stagger} and uses centered differences in $r$ for the Laplacian terms. From Eq.~(\ref{eq:T_axion_def}) one can get
\begin{equation}\label{eq:phi_update}
\partial_t \phi = \frac{c}{2}(T_a^- + T_a^+)~, \qquad \partial_x \phi = \frac{1}{2}(T_a^- - T_a^+)
\end{equation}
then it allows us to update $\phi$ at $(i+1/2, j+1/2)$ by integrating $\partial_t \phi$ over the full step:
\begin{equation}
\phi_{i+1/2,\,j+1/2}^{\,n+1} = \phi_{i+1/2,\,j+1/2}^{\,n} + \frac{c\Delta t}{4}\Big[T_{a,\,i+1,\,j+1/2}^{-,\,n+1/2} + T_{a,\,i,\,j+1/2}^{-,\,n+1/2} + T_{a,\,i+1,\,j+1/2}^{+,\,n+1/2} + T_{a,\,i,\,j+1/2}^{+,\,n+1/2}\Big],
\end{equation}
which completes the axion field update to time $t^{n+1}$. Averaging the face-centered transport variables onto the cell center in this way maintains the second-order accuracy of the scheme.

To model the AREM, we note that the axion-modified Maxwell system for the perturbative fields $(\mathbf{E}_1,\mathbf{B}_1)$ reads~\cite{an_modeling_2024}
\begin{align}
\nabla\times\mathbf{E}_1 &= -\partial_t \mathbf{B}_1~, \\
\nabla\times\mathbf{B}_1 &= \frac{1}{c^2}\partial_t \mathbf{E}_1 + \mu_0 \mathbf{j}_1 + \frac{\gagg}{c}\big[(\partial_t \phi)\mathbf{B}_0 - \mathbf{E}_0\times\nabla \phi\big]~,
\end{align}
where $(\mathbf{E}_0,\mathbf{B}_0)$ denotes the zeroth-order background EM field and $(\mathbf{E}_1,\mathbf{B}_1)$ captures the AREM~\cite{an_modeling_2024}. We evolve this perturbative system with a duplicate QDS solver whose update formulas are identical to Eqs.~\eqref{eq:Trp_update}--\eqref{eq:Ex_update}. The only difference is that the current entering the source terms gains an additional effective contribution
\begin{equation}
\mathbf{j}_a = \frac{\gagg}{c}\big(\mathbf{B}_0\,\partial_t\phi - \mathbf{E}_0\times\nabla\phi\big).
\end{equation}

The axion current is expanded mode by mode with the same convolution operator defined in Eq.~\eqref{eq:mode_convolution}. It can be explicitly written as:
\begin{align}
j_{a,x}^{(m)} &= \frac{\gagg}{c}\Big[\mathcal{C}_m\{B_x,\partial_t\phi\} - \frac{1}{r}\mathcal{C}_m\{E_r,\partial_\theta\phi\} + \mathcal{C}_m\{E_\theta,\partial_r\phi\}\Big], \label{eq:jax_mode}\\
j_{a,r}^{(m)} &= \frac{\gagg}{c}\Big[\mathcal{C}_m\{B_r,\partial_t\phi\} - \mathcal{C}_m\{E_\theta,\partial_x\phi\} + \frac{1}{r}\mathcal{C}_m\{E_x,\partial_\theta\phi\}\Big], \label{eq:jar_mode}\\
j_{a,\theta}^{(m)} &= \frac{\gagg}{c}\Big[\mathcal{C}_m\{B_\theta,\partial_t\phi\} - \mathcal{C}_m\{E_x,\partial_r\phi\} + \mathcal{C}_m\{E_r,\partial_x\phi\}\Big], \label{eq:jat_mode}
\end{align}
where derivatives act on each mode prior to the convolution (e.g. $(\partial_\theta \phi)^{(l)} = -\mi l\,\phi^{(l)}$). These expressions make explicit that $j_{a,\alpha}^{(m)}$ is assembled from all permissible pairs of $\phi^{(l)}$ and the electromagnetic components $E_\beta^{(k)}$, $B_\beta^{(k)}$ with the condition $m=\pm k\pm l$.

\subsection{$r=0$ boundary conditions}
\begin{figure}[!tb]
\centering
\includegraphics[scale=1.0]{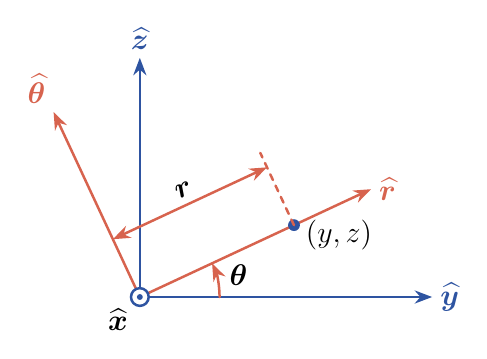}
\caption{\label{fig:coordinates} Coordinate system in Cartesian and cylindrical space.}
\end{figure}
The cylindrical staggering introduces coordinate singularities at the $r=0$ axis, so every field component must satisfy the regularity constraints, as summarized by Lifschitz~\cite{Lifschitz2009}. A physical point at $(r=0,\theta)$ is unique for all $\theta$, which enforces that scalar field $\phi$ and the Cartesian components of any vector field $\mathbf{F}$ must have no $\theta$ dependence at $r = 0$. Thus, the scalar field $\phi$ or the longitudinal component $F_x$ possess only the $m=0$ mode on axis. Consequently, $\phi^{(m)}(r=0)=0$ for $m\ne 0$ and $F_x^{(m)}(r=0)=0$ for $m\ne 0$. The transverse components obey $F_y=F_r\cos\theta-F_\theta\sin\theta$ and $F_z=F_r\sin\theta+F_\theta\cos\theta$, as shown in Fig~\ref{fig:coordinates}. Differentiating with respect to $\theta$ and invoking their Fourier series implies
\begin{equation}
F_r^{(m)} = \frac{\mi}{m} F_\theta^{(m)}, \qquad F_r^{(m)} = \mi m F_\theta^{(m)},
\end{equation}
so $F_r^{(m)}=F_\theta^{(m)}=0$ for $m\ne 1$, while the $m=1$ pair satisfies $F_r^{(1)}=\mi F_\theta^{(1)}$.

Moreover, in our discretization, $E_x$ and $T_\theta^\pm=E_\theta\pm cB_r$ sit exactly at $r_0=0$, whereas $B_\theta$, $B_x$, $T_r^\pm$, and $\phi$ are collocated at half-integer radial offsets $r_{1/2}=\Delta r/2$. We therefore evaluate their on-axis updates by taking the $r\to 0$ limits of the continuous equations and replacing the resulting derivatives by centered differencing stencils. The longitudinal electric field evolves through
\begin{equation}\label{eq:axis_Ex}
\left.\partial_t E_x^{(0)}\right|_{i+1/2,\,0} = 2c^2\left.\partial_r B_\theta^{(0)}\right|_{i+1/2,\,0} - \frac{1}{\varepsilon_0}j_{x,\,i+1/2,\,0}^{(0)},
\end{equation}
where the derivative is approximated by
\begin{equation}
\left.\partial_r B_\theta^{(0)}\right|_{i+1/2,\,0} \approx \frac{2}{\Delta r} B_{\theta,\,i+1/2,\,1/2}^{(0)}.
\end{equation}
Similarly, the on-axis transport variables obey
\begin{equation}\label{eq:axis_Ttheta}
\left.\partial_t T_\theta^{\pm,(1)}\right|_{i+1/2,\,0} \mp c\left.\partial_x T_\theta^{\pm,(1)}\right|_{i+1/2,\,0} = -c^2\left.\partial_r B_x^{(1)}\right|_{i+1/2,\,0} - \frac{1}{\varepsilon_0}j_{\theta,\,i+1/2,\,0}^{(1)} \pm \mi c\left.\partial_r E_x^{(1)}\right|_{i+1/2,\,0},
\end{equation}
with the required radial derivatives evaluated as
\begin{equation}
\left.\partial_r B_x^{(1)}\right|_{i+1/2,\,0} \approx \frac{2}{\Delta r} B_{x,\,i+1/2,\,1/2}^{(1)}, \qquad \left.\partial_r E_x^{(1)}\right|_{i+1/2,\,0} \approx \frac{E_{x,\,i+1/2,\,1}^{(1)} - E_{x,\,i+1/2,\,0}^{(1)}}{\Delta r}.
\end{equation}
Because $E_x^{(1)}(r=0)=0$, the second expression reduces to $E_{x,\,i+1/2,\,1}^{(1)}/\Delta r$ in practice. Equation~\eqref{eq:axis_Ttheta} furnishes the axis update for $T_\theta^\pm$ without introducing singular denominators.

The relations $E_r^{(1)}=\mi E_\theta^{(1)}$ and $B_r^{(1)}=\mi B_\theta^{(1)}$ imply a coupling between the on-axis radial and azimuthal transport variables, yielding
\begin{equation}\label{eq:axis_Tr}
T_{r,\,i+1,\,0}^{\pm,(1)} = \mi T_{\theta,\,i+1,\,0}^{\mp,(1)}.
\end{equation}
We impose Eq.~\eqref{eq:axis_Tr} by reflecting $T_r^\pm$ across the axis with a ghost cell $j=-1/2$:
\begin{equation}
T_{r,\,i+1,\,-1/2}^{\pm,(1)} = 2\mi T_{\theta,\,i+1,\,0}^{\mp,(1)} - T_{r,\,i+1,\,1/2}^{\pm,(1)},
\end{equation}
which maintains the second-order accuracy of the discrete stencil. For $B_x$ and $\phi$ we enforce the homogeneous Neumann conditions,
\begin{equation}
\left.\partial_r B_x^{(m)}\right|_{i+1/2,\,0} = 0, \qquad \left.\partial_r \phi^{(m)}\right|_{i+1/2,\,0} = 0,
\end{equation}
implemented by mirroring the first off-axis cell values, e.g. $B_{x,\,i+1/2,\,-1/2}^{(m)} = B_{x,\,i+1/2,\,1/2}^{(m)}$ and $\phi_{\,i+1/2,\,-1/2}^{(m)} = \phi_{\,i+1/2,\,1/2}^{(m)}$.

\section{PIC implementation and boundary treatment}\label{sec:implementation}
\subsection{Particle advance and charge-conserving current deposition}
The QDS field solver is coupled to the particle and current routines of cylindrical EPOCH~\cite{Arber2015,cylindrical_epoch_github}. A macroparticle is represented by the Cartesian coordinates $(x_p,y_p,z_p)$ and normalized momentum $\bm{u}_p=\bm{p}_p/(m_pc)$, while the fields are stored on the $(x,r)$ mesh as azimuthal modes. At the particle position, $r_p=(y_p^2+z_p^2)^{1/2}$ and $\theta_p=\operatorname{atan2}(z_p,y_p)$. A cylindrical field component is interpolated according to
\begin{equation}\label{eq:particle_field_gather}
F_p=\sum_{a,b}g_{x,i+a}(x_p)g_{r,j+b}(r_p)
\text{Re}\!\left\{\sum_{m=0}^{m_{\max}}F_{i+a,j+b}^{(m)}e^{-\mi m\theta_p}\right\},
\end{equation}
where the indices $(i+a,j+b)$ follow the staggered position of the corresponding field component, and $a$, $b$ represent the neighbour cells contributed to the components. The factor $g_{\alpha,\ell}$ is the one-dimensional discrete shape weight along $\alpha\in\{x,r\}$, evaluated on the grid associated with that field component. Equivalently, it is the cell-integrated weight of the underlying particle shape $S_\alpha$ over the interval centered at the corresponding staggered grid point. Thus Eq.~\eqref{eq:particle_field_gather} uses the same discrete shape weights as the current deposition. The simulations presented below use the triangular shape function. After interpolation, the cylindrical components are transformed as $E_y=E_r\cos\theta_p-E_\theta\sin\theta_p$ and $E_z=E_r\sin\theta_p+E_\theta\cos\theta_p$, with the same transformation for the magnetic field. The QDS magnetic fields are centered onto the field positions used by the original particle interpolation before Eq.~\eqref{eq:particle_field_gather} is applied. This centering is incorporated into the interpolation weights and leaves the particle shape and interpolation order unchanged.

The particle motion is advanced with the relativistic Boris method~\cite{Boris1970}. The particle is first drifted by half a time step to the position at which the fields in Eq.~\eqref{eq:particle_field_gather} are evaluated. With $\bm{u}_p^n$ denoting the momentum before the force update, the two electric accelerations and the magnetic rotation are
\begin{align}
\bm{u}^{-}_p&=\bm{u}_p^n+\frac{q_p\Delta t}{2m_pc}\bm{E}_p,\qquad
\bm{t}_p=\frac{q_p\Delta t}{2m_p\gamma_p^{-}}\bm{B}_p,
\qquad \bm{s}_p=\frac{2\bm{t}_p}{1+|\bm{t}_p|^2},\label{eq:boris_parameters}\\
\bm{u}'_p&=\bm{u}^{-}_p+\bm{u}^{-}_p\times\bm{t}_p,\qquad
\bm{u}^{+}_p=\bm{u}^{-}_p+\bm{u}'_p\times\bm{s}_p,\qquad
\bm{u}_p^{n+1}=\bm{u}^{+}_p+\frac{q_p\Delta t}{2m_pc}\bm{E}_p,\label{eq:boris_update}
\end{align}
where $\gamma_p^{-}=(1+|\bm{u}_p^{-}|^2)^{1/2}$. A second half-step drift with $\bm{v}_p^{n+1}=c\bm{u}_p^{n+1}/\sqrt{1+|\bm{u}_p^{n+1}|^2}$ completes the position update. Therefore, the QDS solver changes only the field values supplied to the particle interpolation. The particle equations and the Boris update are identical to those of cylindrical EPOCH.

The current is deposited with the charge-conserving method developed for cylindrical EPOCH from the Esirkepov construction~\cite{Esirkepov2001}. Its basic principle is to equate the change of the charge carried by a particle shape in each cell to the current through the cell faces during the same step. Let $Q_p$ be the charge represented by a macroparticle. In the deposition derivation, $g_{x,i}^{0}$ and $g_{r,j}^{0}$ denote the old values of the same discrete shape weights on the deposition grid before the push. After the push, these weights are $g_{x,i}^{1}=g_{x,i}^{0}+h_{x,i}$ and $g_{r,j}^{1}=g_{r,j}^{0}+h_{r,j}$, where $h_{x,i}$ and $h_{r,j}$ are the corresponding weight changes during the push. We parameterize the particle trajectory over one time step by $\xi\in[0,1]$,
\begin{equation}\label{eq:particle_deposition_path}
g_{x,i}(\xi)=g_{x,i}^{0}+\xi h_{x,i},\qquad
g_{r,j}(\xi)=g_{r,j}^{0}+\xi h_{r,j},\qquad
\theta_p(\xi)=\theta_p^0+\xi\Delta\theta_p.
\end{equation}
For the field convention in Eq.~\eqref{eq:fourier_decomp}, the contribution of a particle at angle $\theta$ to mode $m$ is weighted by $C_m(\theta)$. It is convenient to introduce the trajectory-average operator
\begin{equation}\label{eq:particle_mode_weight}
C_m(\theta)=
\begin{cases}1,&m=0,\\2e^{\mi m\theta},&m>0,\end{cases}
\qquad
\left\langle W\right\rangle_m=\int_0^1W(\xi)C_m(\theta_p^0+\xi\Delta\theta_p)\,\mathrm{d}\xi .
\end{equation}
Here $W(\xi)$ denotes whichever transverse shape weight is to be averaged along the trajectory. For example, $W=g_{r,j}(\xi)$ for an $x$-directed flux, $W=g_{x,i}(\xi)$ for a radial flux, and $W=g_{x,i}(\xi)g_{r,j}(\xi)$ for the azimuthal current.

The net modal currents associated with the changes $h_{x,i}$ and $h_{r,j}$ are then
\begin{equation}\label{eq:particle_modal_currents}
I_{x,ij}^{(m)}=\frac{Q_p}{\Delta t}h_{x,i}\left\langle g_{r,j}^{0}+\xi h_{r,j}\right\rangle_m,
\qquad
I_{r,ij}^{(m)}=\frac{Q_p}{\Delta t}h_{r,j}\left\langle g_{x,i}^{0}+\xi h_{x,i}\right\rangle_m,
\end{equation}
and
\begin{equation}\label{eq:particle_azimuthal_current}
j_{\theta,ij}^{(m)}=\frac{Q_pv_{\theta,p}}{V_{ij}}
\left\langle(g_{x,i}^{0}+\xi h_{x,i})(g_{r,j}^{0}+\xi h_{r,j})\right\rangle_m .
\end{equation}
Here $V_{ij}=2\pi r_j\Delta x\Delta r$ is the cylindrical cell volume and $v_{\theta,p}$ is evaluated from the Cartesian velocity at the temporal midpoint of the deposition trajectory. In the implementation, the integrals in $\xi$ are evaluated analytically. For $|m\Delta\theta_p|<10^{-4}$, the corresponding series expansions are used to avoid cancellation in terms containing $e^{\mi m\Delta\theta_p}-1$.

Charge conservation follows directly from using the same trajectory in the charge and current weights. The modal charge carried by this particle in cell $(i,j)$ is proportional to
\begin{equation}\label{eq:particle_modal_charge}
q_{ij}^{(m)}(\xi)=Q_p g_{x,i}(\xi)g_{r,j}(\xi)C_m[\theta_p(\xi)].
\end{equation}
Its change is indeed the difference between Eq.~\eqref{eq:particle_modal_charge} evaluated at the two endpoints:
\begin{align}
\Delta q_{ij}^{(m)}
=\int_0^1\frac{\mathrm{d} q_{ij}^{(m)}}{\mathrm{d}\xi}\,\mathrm{d}\xi
= {}&Q_p h_{x,i}\left\langle g_{r,j}\right\rangle_m
+Q_p h_{r,j}\left\langle g_{x,i}\right\rangle_m\nonumber\\
&+\mi m Q_p\Delta\theta_p\left\langle g_{x,i}g_{r,j}\right\rangle_m .\label{eq:particle_charge_identity}
\end{align}
Here $\mathrm{d}g_{x,i}/\mathrm{d}\xi=h_{x,i}$, $\mathrm{d}g_{r,j}/\mathrm{d}\xi=h_{r,j}$, and $\mathrm{d}C_m/\mathrm{d}\xi=\mi m\Delta\theta_p C_m$ for $m>0$ ($\mathrm{d}C_m/\mathrm{d}\xi=0$ for $m=0$). The first two terms in Eq.~\eqref{eq:particle_charge_identity} are $\Delta t I_{x,ij}^{(m)}$ and $\Delta t I_{r,ij}^{(m)}$, while the last is the modal charge change due to azimuthal motion.

The quantities $I_x^{(m)}$ and $I_r^{(m)}$ determine the net current required in a cell, but do not by themselves specify the current density on either face. Let $A_{r\theta,j}=2\pi r_j\Delta r$ be the area normal to $x$, and $A_{x\theta,j\pm1/2}=2\pi r_{j\pm1/2}\Delta x$ the radial-face areas. The face currents are chosen so that their net outward flux reproduces the negative of the shape-change currents,
\begin{align}
A_{r\theta,j}\left(j_{x,i+1/2,j}^{(m)}-j_{x,i-1/2,j}^{(m)}\right)&=-I_{x,ij}^{(m)},\label{eq:face_current_x}\\
A_{x\theta,j+1/2}j_{r,i,j+1/2}^{(m)}-A_{x\theta,j-1/2}j_{r,i,j-1/2}^{(m)}&=-I_{r,ij}^{(m)}.\label{eq:face_current_r}
\end{align}

Dividing Eqs.~\eqref{eq:face_current_x} and \eqref{eq:face_current_r} by the cell volume $V_{ij}$ converts the first two terms of Eq.~\eqref{eq:particle_charge_identity} into the axial and radial parts of $-\nabla\cdot\mathbf{j}$. For the angular motion, $\dot\theta_p=\Delta\theta_p/\Delta t=v_{\theta,p}/r_p$ at the midpoint accuracy of the particle push. Since Eq.~\eqref{eq:particle_azimuthal_current} deposits the azimuthal charge flux with the same full shape average, its Fourier-space divergence supplies the corresponding single-particle contribution
\begin{equation}\label{eq:angular_current_cancellation}
-\frac{\mi m}{r_j}j_{\theta,ij,p}^{(m)}
=
-\frac{\mi m Q_p\Delta\theta_p}{V_{ij}\Delta t}
\left\langle g_{x,i}g_{r,j}\right\rangle_m,
\end{equation}
which is the negative of the third term in Eq.~\eqref{eq:particle_charge_identity} after division by $V_{ij}\Delta t$. Therefore, after summing the single-particle identity over all particles, one obtains, for every retained mode,
\begin{align}
\frac{\rho_{i,j}^{(m),n+1}-\rho_{i,j}^{(m),n}}{\Delta t}
&+\frac{j_{x,i+1/2,j}^{(m),n+1/2}-j_{x,i-1/2,j}^{(m),n+1/2}}{\Delta x}\nonumber\\
&+\frac{r_{j+1/2}j_{r,i,j+1/2}^{(m),n+1/2}-r_{j-1/2}j_{r,i,j-1/2}^{(m),n+1/2}}{r_j\Delta r}
-\frac{\mi m}{r_j}j_{\theta,i,j}^{(m),n+1/2}=0.\label{eq:discrete_continuity}
\end{align}
Thus the deposited current satisfies the discrete continuity equation by construction. The QDS field update uses these currents at the same staggered positions as the original cylindrical solver. When a current is required at an intermediate position in Eqs.~\eqref{eq:Trp_update}--\eqref{eq:Ttm_update}, the two neighboring values are averaged. This field-solver interpolation does not alter the charge-conserving deposition itself.

\subsection{Open field boundaries}
At the axial boundaries, a source-free open boundary is imposed by setting the inward-propagating transport variables to zero,
\begin{equation}\label{eq:em_x_open}
T_r^+=T_\theta^-=0\quad (x=x_{\min}),\qquad
T_r^-=T_\theta^+=0\quad (x=x_{\max}),
\end{equation}
while the outward-propagating variables are updated by Eqs.~\eqref{eq:Trp_update}--\eqref{eq:Ttm_update}. The corresponding conditions for the axion transport variables are
\begin{equation}\label{eq:axion_x_open}
T_a^+=0\quad (x=x_{\min}),\qquad T_a^-=0\quad (x=x_{\max}).
\end{equation}

At $r=R_{\max}$, the electromagnetic field is assumed to be locally transverse and outward propagating, as in the outflow boundary of cylindrical EPOCH,
\begin{equation}\label{eq:em_r_characteristic}
E_\theta^{(m)}=cB_x^{(m)},\qquad E_x^{(m)}=-cB_\theta^{(m)}.
\end{equation}
Combining these relations with Maxwell equations gives
\begin{align}
(\partial_t+c\partial_r)B_x^{(m)}
&=\frac{c}{2}\partial_x B_r^{(m)}
-\frac{E_\theta^{(m)}+\mi m E_r^{(m)}}{2r}
-\frac{j_\theta^{(m)}}{2\varepsilon_0c},\label{eq:em_r_open_bx}\\
(\partial_t+c\partial_r)B_\theta^{(m)}
&=-\frac{1}{2}\partial_x E_r^{(m)}
-\frac{c}{2r}\left(B_\theta^{(m)}+\mi m B_r^{(m)}\right)
+\frac{j_x^{(m)}}{2\varepsilon_0c}.\label{eq:em_r_open_bt}
\end{align}
At the outermost radial point, Eqs.~\eqref{eq:em_r_open_bx} and \eqref{eq:em_r_open_bt} are centered in time and solved explicitly for $B_x^{(m)}$ and $B_\theta^{(m)}$. The quantities on the right-hand sides are averaged to the positions of the corresponding boundary fields according to the staggering in Fig.~\ref{fig:stagger}.

For the axion field, the outer-radial boundary is obtained by combining the outgoing condition
\begin{equation}\label{eq:axion_r_sommerfeld}
\left(\partial_t+c\partial_r\right)\phi^{(m)}=0
\qquad (r=R_{\max})
\end{equation}
with the mode-decomposed Klein--Gordon equation. Let $j=N_r$ denote the outermost radial field point, $r_{N_r}=R_{\max}$, and $j=N_r+1$ the exterior ghost point. Centering Eq.~\eqref{eq:axion_r_sommerfeld} at $(r_{N_r},t^n)$ gives
\begin{equation}\label{eq:axion_r_ghost}
\phi_{i,N_r+1}^{(m),n}=\phi_{i,N_r-1}^{(m),n}
-\frac{\Delta r}{c\Delta t}\left(\phi_{i,N_r}^{(m),n+1}-\phi_{i,N_r}^{(m),n-1}\right).
\end{equation}
Substitution of Eq.~\eqref{eq:axion_r_ghost} into the centered Klein--Gordon equation eliminates the exterior value and yields an explicit boundary update,
\begin{align}
A_{N_r}\phi_{i,N_r}^{(m),n+1}
={}&-\left(\frac{1}{\Delta t^2}-\frac{c}{\Delta r\Delta t}-\frac{c}{2r_{N_r}\Delta t}\right)
\phi_{i,N_r}^{(m),n-1}
+\frac{2}{\Delta t^2}\phi_{i,N_r}^{(m),n}
+\mathcal{R}_{i,N_r}^{(m),n},\label{eq:axion_r_update}\\
A_{N_r}={}&\frac{1}{\Delta t^2}+\frac{c}{\Delta r\Delta t}+\frac{c}{2r_{N_r}\Delta t},\nonumber\\
\mathcal{R}_{i,N_r}^{(m),n}={}&c^2\frac{\phi_{i+1,N_r}^{(m),n}-2\phi_{i,N_r}^{(m),n}+\phi_{i-1,N_r}^{(m),n}}{\Delta x^2}
-\frac{m^2c^2}{r_{N_r}^2}\phi_{i,N_r}^{(m),n}
+\frac{2c^2}{\Delta r^2}\left(\phi_{i,N_r-1}^{(m),n}-\phi_{i,N_r}^{(m),n}\right)\nonumber\\
&-\frac{m_a^2c^4}{\hbar^2}\phi_{i,N_r}^{(m),n}
+\frac{\gagg c^2}{\hbar\mu_0}S_{a,i,N_r}^{(m),n}.\nonumber
\end{align}
After Eq.~\eqref{eq:axion_r_update} is evaluated, Eq.~\eqref{eq:axion_r_ghost} gives the exterior ghost-cell value. The same update is applied to the axion field at the auxiliary half time level.

\subsection{MPI domain decomposition}
The $(x,r)$ mesh is divided into an $n_x^{\rm MPI}\times n_r^{\rm MPI}$ array of rectangular subdomains. Each MPI process stores all retained azimuthal modes in its local subdomain. Field values in the ghost cells are exchanged between neighboring subdomains, and particles crossing an internal boundary are transferred to the neighboring process. The QDS transport variables require the same nearest-neighbor exchange because their axial update spans one cell and their radial derivatives use centered differences. No global field operation is introduced by the QDS solver.

For future simulations requiring a large number of azimuthal modes, the evaluation of $S_a=\mathbf{E}\cdot\mathbf{B}$ may be accelerated by temporarily transforming the electromagnetic fields from Fourier space to a discrete $\theta$ grid, evaluating the scalar product pointwise, and transforming the result back to obtain $S_a^{(m)}$. This procedure replaces the direct mode convolution in Eq.~\eqref{eq:mode_convolution} by a pair of FFT-based transforms. Since the domain is decomposed only in the $(x,r)$ directions and every process retains the complete set of azimuthal modes in its local subdomain, these transforms can be performed independently by each process without additional inter-process communication. The direct convolution remains preferable for a small number of retained modes. The FFT-based procedure becomes advantageous only when its transform cost is lower than that of evaluating the mode convolution explicitly.

\section{Benchmark Results}\label{sec:benchmarks}

\subsection{Laser propagation in vacuum}

We first isolate the numerical dispersion of the electromagnetic solver by propagating a laser pulse in vacuum. The benchmark uses a circularly polarized Gaussian beam propagating along $x$, with wavelength $\lambda_0=800\,\mathrm{nm}$ and normalized amplitude $a_0=5$, where $E_0\equiv |m_e\omega_0 c/e|$. The transverse waist and longitudinal pulse length are both set to $10\lambda_0$, corresponding to a temporal duration parameter $\tau_L=10T_0$ with $T_0=2\pi/\omega_0$. No plasma particles are included.

For the quasi-cylindrical QDS and Yee simulations, the initial domain is $-40\lambda_0\le x\le 40\lambda_0$ and $0\le r\le 40\lambda_0$. Four azimuthal modes are retained ($m\le 3$). The reference low-resolution runs use $\Delta x=\lambda_0/10$ and $\Delta r=\lambda_0/2$. The QDS run uses the special time step $c\Delta t=\Delta x$, whereas the standard Yee run uses the stable explicit time step corresponding to $\Delta t=0.08T_0$. We also include the Lehe finite-difference solver implemented in EPOCH~\cite{Lehe2013} as a Cartesian 3D comparison, using $\Delta x=\lambda_0/10$, and $\Delta y=\Delta z=\lambda_0/2$. For the QDS resolution scan, the axial resolution is varied from $\Delta x=\lambda_0/10$ to $\lambda_0/40$ while keeping $\Delta r=\lambda_0/2$.

The group velocity diagnostic is based on the motion of the pulse energy centroid. At each output time we integrate the transverse electromagnetic energy density over the transverse plane to obtain an axial profile,
\begin{equation}
    U(x,t)=\int \frac{1}{2}\left(\varepsilon_0 |\mathbf{E}_\perp|^2+\frac{1}{\mu_0}|\mathbf{B}_\perp|^2\right)\,\mathrm{d}A,
\end{equation}
where $\mathrm{d}A=2\pi r\,\mathrm{d}r$ in the quasi-cylindrical simulations and $\mathrm{d}A=\mathrm{d}y\,\mathrm{d}z$ in the Cartesian Lehe run. The centroid position is then
\begin{equation}
    x_c(t)=\frac{\int x U(x,t)\,\mathrm{d}x}{\int U(x,t)\,\mathrm{d}x},
\end{equation}
and the measured group velocity is obtained from a local linear fit, $v_g=\mathrm{d}x_c/\mathrm{d}t$.

For a paraxial Gaussian beam in vacuum, the reduction of the axial group velocity arises from the finite transverse wave-vector content~\cite{Giovannini2015}. To leading order,
\begin{equation}
    \frac{v_g}{c}\simeq 1-\frac{\langle k_\perp^2\rangle}{2k_0^2}.
\end{equation}
For the energy-weighted Gaussian profile used here, $\langle k_\perp^2\rangle=2/w_0^2$, giving the theoretical reference
\begin{equation}\label{eq:vacuum_gaussian_group_velocity}
    \frac{v_g}{c}\simeq 1-\left(\frac{\lambda_0}{2\pi w_0}\right)^2.
\end{equation}
With $w_0=10\lambda_0$, this yields $|v_g/c-1|=2.53\times10^{-4}$.

\begin{figure}[!tb]
    \centering
    \includegraphics[width = 0.6\textwidth]{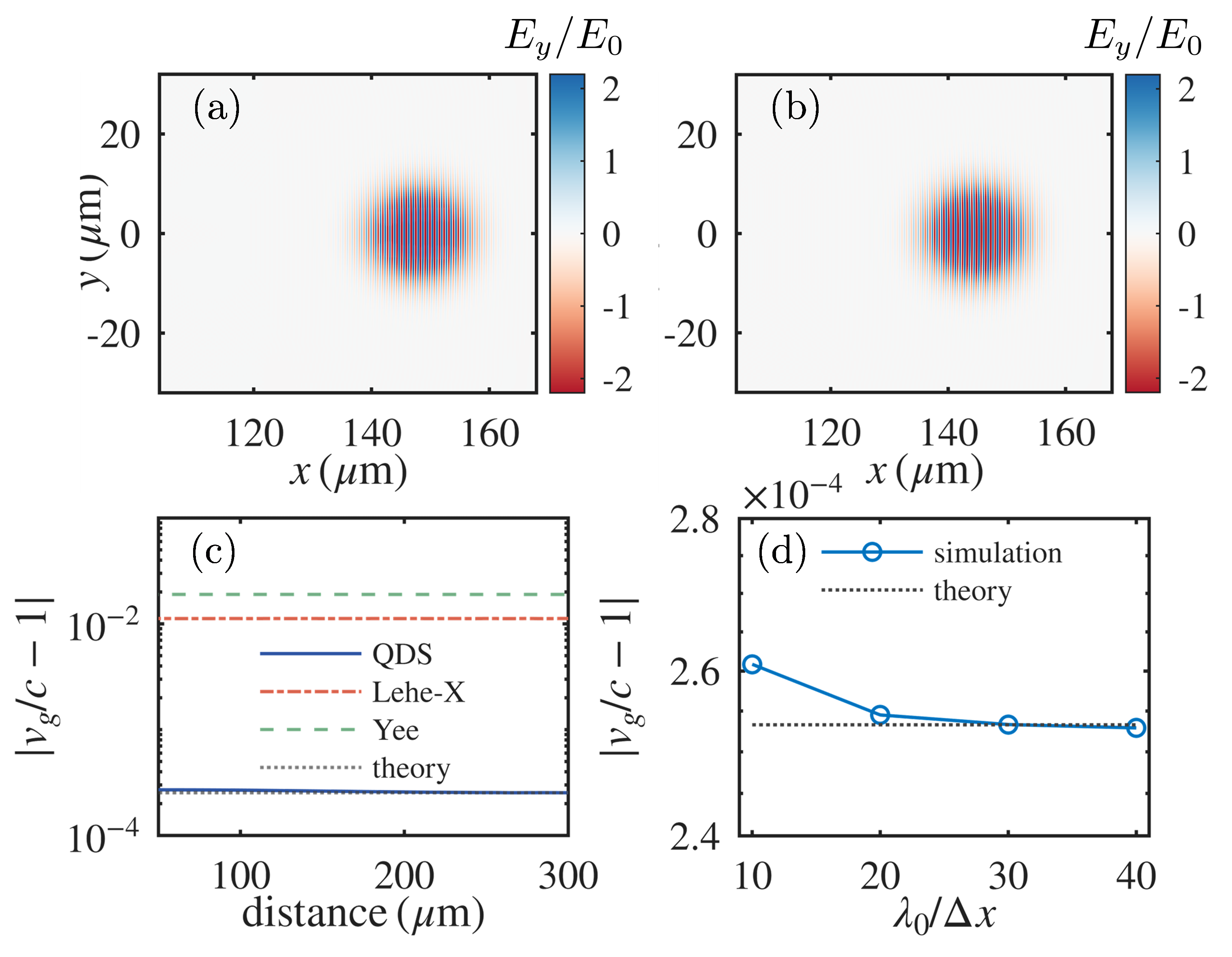}
    \caption{Benchmark of laser propagation in vacuum. (a) Electric field $E_y$ from QDS solver. (b) Electric field $E_y$ from standard Yee solver. (c) Comparison of the energy-centroid group velocity of the laser pulse between different solvers, where the QDS result nearly overlaps with the theoretical reference. The Lehe solver~\cite{Lehe2013} is the 3D Cartesian implementation in EPOCH. (d) Resolution scan of the QDS solver, where only the axial resolution is varied.}
    \label{fig:vacuum_propagation}
\end{figure}

Figure~\ref{fig:vacuum_propagation} shows that the QDS solver preserves the expected vacuum propagation of the laser pulse. At $\Delta x=\lambda_0/10$, the QDS result gives $|v_g/c-1|\simeq2.61\times10^{-4}$ over the propagation interval $50$--$300\,\mu\mathrm{m}$, already very close to Eq.~\eqref{eq:vacuum_gaussian_group_velocity}. By contrast, the standard Yee solver strongly underestimates the group velocity at this resolution. The Lehe solver reduces the error relative to Yee, as expected from its modified finite-difference stencil, but with the resolution parameters used here it still introduces a visible artificial change of the laser velocity.

The QDS convergence with axial resolution is shown in Fig.~\ref{fig:vacuum_propagation}(d). Increasing the resolution from $\Delta x=\lambda_0/10$ to $\lambda_0/40$ changes the measured $|v_g/c-1|$ from $2.61\times10^{-4}$ to $2.53\times10^{-4}$. Thus even the lowest-resolution QDS run captures the physical group velocity to high accuracy, and further increasing the axial resolution produces only a small refinement toward the theoretical value.

\subsection{Laser wakefield excitation}
To verify the accuracy and efficiency of the new algorithm in plasma, we compare the cylindrical QDS solver, the cylindrical standard Yee solver, and a fully 3D Cartesian simulation. The laser parameters are the same as those in the vacuum-propagation benchmark. The plasma has a plateau density $n_0=0.001\,n_c\approx 1.7\times 10^{18}\,\mathrm{cm}^{-3}$ preceded by a $40\lambda_0$ linear up-ramp, where $n_c\equiv m_e\varepsilon_0\omega_0^2/e^2$ is the critical density. The cylindrical QDS and Yee runs also use the same low-resolution quasi-cylindrical mesh as in the previous subsection, namely $\Delta x=\lambda_0/10$, $\Delta r=\lambda_0/2$, and $m\le 3$. The fully 3D run uses $\Delta x=\lambda_0/20$ and $\Delta y=\Delta z=\lambda_0/2$. The time steps are $c\Delta t=\Delta x$ for QDS, $\Delta t=0.08T_0$ for the cylindrical Yee solver, and $\Delta t=0.04T_0$ for the 3D run.

\begin{figure}[!tb]
    \centering
    \includegraphics[width = 0.84\textwidth]{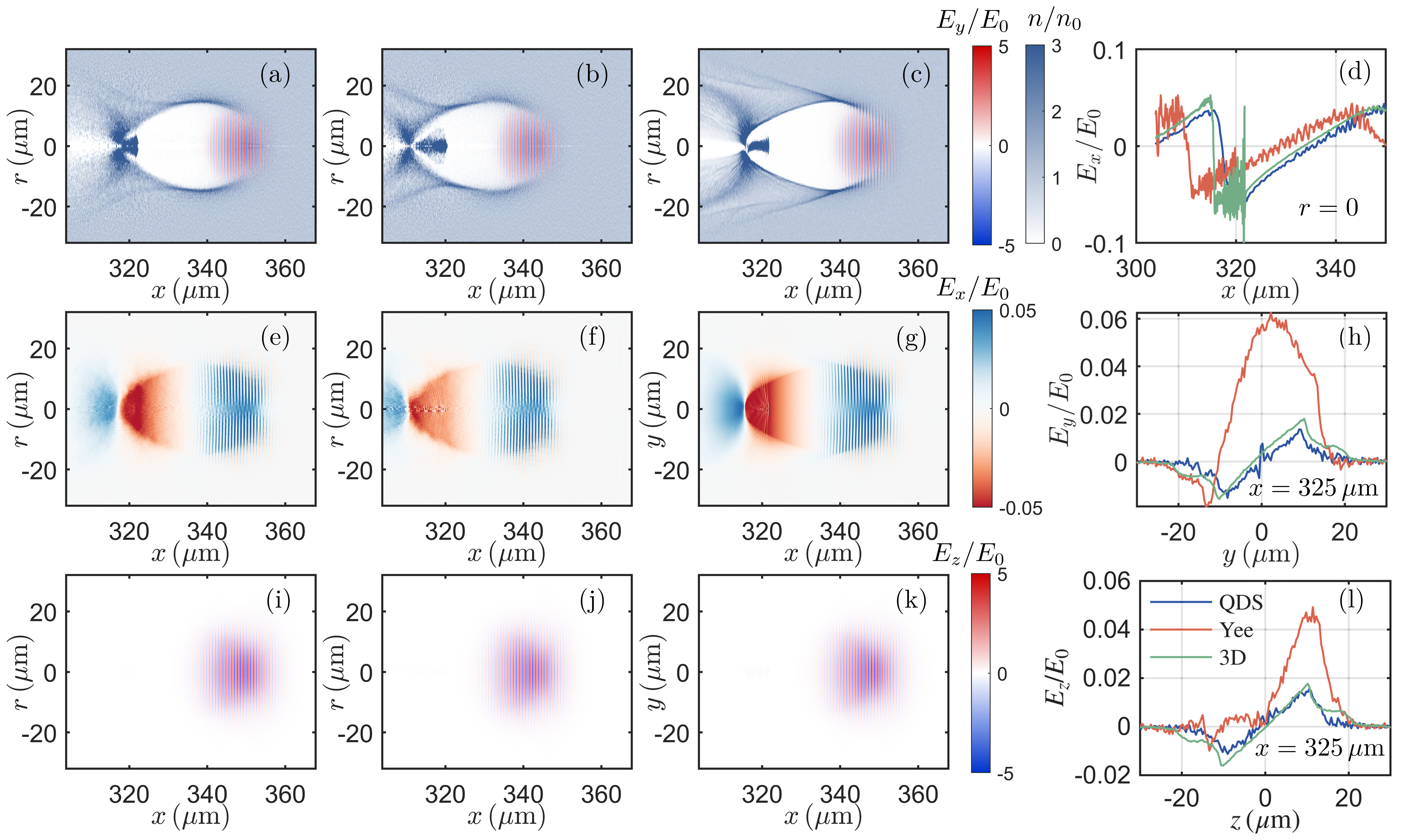}
    \caption{Laser wakefield excitation benchmark. The first three columns show the field distributions in the $x$-$y$ plane from the cylindrical QDS solver (1st column), the cylindrical standard Yee solver (2nd column), and the fully 3D simulation (3rd column). The last column shows 1D line cuts comparing the three solvers.
    (a)--(d) Longitudinal electric field $E_x$ (overlaid with electron density in gray). (a)--(c) show 2D distributions, and (d) shows the on-axis line cut at $y=0, z=0$.
    (e)--(h) Transverse electric field $E_y$. (e)--(g) show 2D distributions, and (h) shows the transverse line cut at $x=325\,\mu\text{m}$.
    (i)--(l) Transverse electric field $E_z$. (i)--(k) show 2D distributions, and (l) shows the transverse line cut at $x=325\,\mu\text{m}$.}
    \label{fig:wakefield}
\end{figure}

Figure~\ref{fig:wakefield} compares the wakefield structure and the laser-driven bubble among the three calculations. The field distributions in the first three columns show consistent bubble shapes across all cases, indicating that the cylindrical implementation reproduces the expected wake excitation. The 1D line cuts [Figs.~\ref{fig:wakefield}(d), (h), and (l)] provide a quantitative comparison. In the standard Yee solver, both the pulse and the wakefield are slightly delayed due to numerical dispersion. The higher-resolution 3D run mitigates this dispersion to some extent, but a small residual delay remains.

The transverse-field line cuts in Figs.~\ref{fig:wakefield}(h) and (l) show larger deviations for the Yee calculation than for the other two solvers. This difference is a direct consequence of the different field evolution in the simulations, but it should not be interpreted as a pointwise error of the wakefield alone. Because the Yee laser pulse has a stronger numerical dispersion and therefore a lower group velocity, it is shifted backward relative to the QDS and 3D pulses. At the fixed sampling position used for the line cuts, the transverse field in the Yee run consequently contains a contribution from the laser tail in addition to the plasma wakefield.

In addition to accuracy, the computational cost differs significantly. On the same 80 CPU cores, the QDS run takes about 10 minutes, the cylindrical Yee run about 20 minutes, whereas the fully 3D run takes about 63.5 hours for the same physical scenario. This shows the accuracy and efficiency advantages of the cylindrical QDS solver for laser-plasma interaction problems.

\begin{figure}[!tb]
    \centering
    \includegraphics[width=0.74\textwidth]{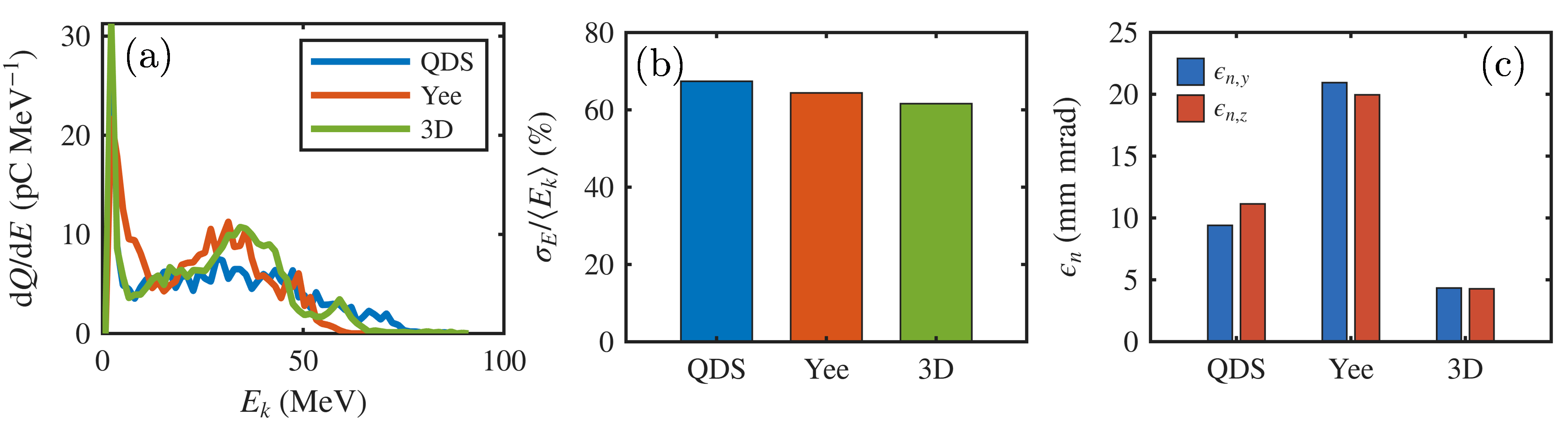}
    \caption{Electron beam accelerated from the wakefield. (a) Electron beam energy spectrum from different solvers. (b) Comparison of the electron beam energy spread. (c) Comparison of the electron beam emittance.}
    \label{fig:lwfa_electron_beam}
\end{figure}

Figure~\ref{fig:lwfa_electron_beam} further compares the electron beams accelerated in the three runs. The low-energy tail of the spectrum from the QDS solver is closer to the 3D result, whereas the Yee solver produces more low-energy injected electrons. This trend is consistent with the stronger numerical dispersion of the Yee solver, which reduces the laser group velocity and enhances the self-injection. Moreover, all the three cases give similar electron beam energy spread. The high-energy spectral shape and the emittance do not coincide exactly between the QDS and 3D runs, but they are still similar. They show that the QDS solver gives the beam spectrum, energy spread, and emittance in a reasonable range at a much lower computational cost, while avoiding the enhanced numerical-dispersion effect observed in the Yee calculation.

\subsection{Resonant axion generation via two-color laser mixing}

To test the solver's capability in handling dispersive wave propagation and phase-sensitive coupling, we simulate axion generation via the interaction of two co-propagating laser pulses in a plasma.

Consider two linearly polarized laser pulses, labeled $A$ and $B$, with angular frequencies $\omega_A$ and $\omega_B$ propagating along the $x$-axis. Pulse $A$ is polarized along $\hat{y}$ ($E_{y,A}, B_{z,A}$) and pulse $B$ along $\hat{z}$ ($E_{z,B}, B_{y,B}$). The axion source term $(S_a \propto \mathbf{E}\cdot\mathbf{B})$ is given by:
\begin{equation}
    \mathbf{E}\cdot\mathbf{B} = E_{y,A} B_{y,B} + E_{z,B} B_{z,A}.
\end{equation}
In a continuous medium, the magnetic fields are related to the electric fields via the phase velocities $v_{pA}$ and $v_{pB}$ as $B_{z,A} = E_{y,A}/v_{pA}$ and $B_{y,B} = -E_{z,B}/v_{pB}$ (assuming forward propagation). The source term thus becomes:
\begin{equation}\label{eq:mixing_source}
    \mathbf{E}\cdot\mathbf{B} = E_{y,A} E_{z,B} \left( \frac{1}{v_{pA}} - \frac{1}{v_{pB}} \right).
\end{equation}
This expression reveals two critical physical features. First, in a vacuum where $v_{pA}=v_{pB}=c$, the source term must vanish identically. Any non-zero signal in a simulation implies a numerical error where the solver assigns different numerical phase velocities to different frequencies. Second, in a plasma, the physical dispersion relation $\omega^2 = \omega_p^2 + c^2k^2$ yields different phase velocities for different frequencies, generating a net source. Efficient axion generation then requires the phase-matching condition $k_a = k_A + k_B$ and energy conservation $\omega_a = \omega_A + \omega_B$, where $k_{A,B}$ and $k_a$ are the wavenumbers of the lasers and the massive axion, respectively.

We simulate this process as a benchmark of the QDS solver. Pulse $A$ has a fundamental wavelength $\lambda_A = 800\,\mathrm{nm}=\lambda_0$ ($\omega_A = \omega_0$), and pulse $B$ is the second harmonic with $\lambda_B = 400\,\mathrm{nm}$ ($\omega_B = 2\omega_0$). Both pulses have a normalized amplitude $a_0 =E_{A}/E_0=E_{B}/E_0= 0.01$ and focus to a Gaussian spot size $w_0 = 40\lambda_0$. They co-propagate into a uniform plasma with density $n_e = 0.01 n_c$ starting from $x=0$. The axion mass $m_a$ is chosen to satisfy the resonance condition $(\omega_A+\omega_B)^2/c^2 = k_a^2 + m_a^2 c^2/\hbar^2$. The simulation resolutions remain consistent with the previous benchmarks ($c\Delta t = \Delta x = \lambda_0/10$). The lasers enter into the simulation box at $x = -64\,\mu\text{m}$ with a delay of 213\,fs.

\begin{figure}[!tb]
    \centering
    \includegraphics[width = 0.86\textwidth]{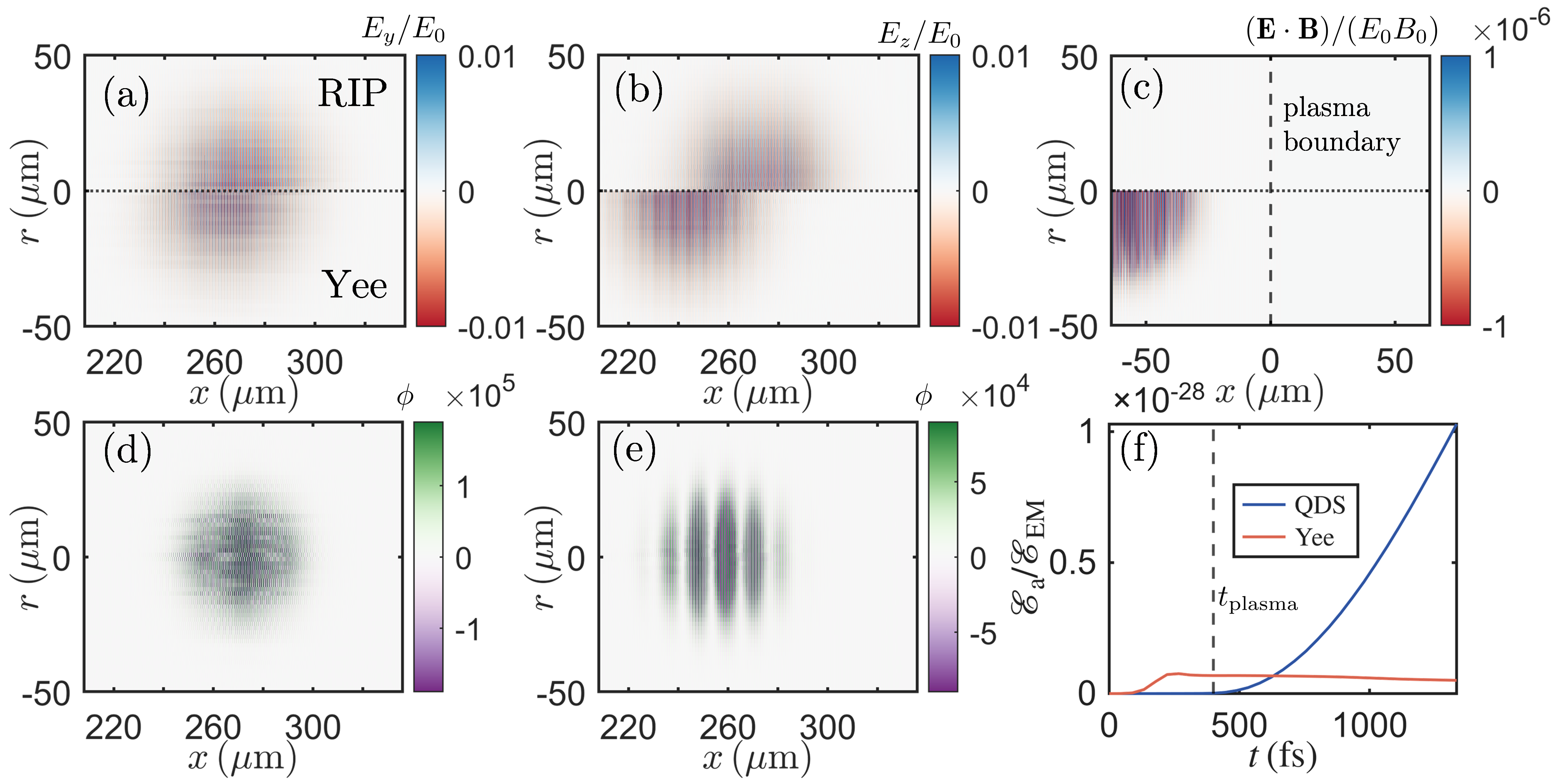}
    \caption{Simulations of resonant axion generation via two-color laser mixing. (a)--(c) show comparisons between the QDS solver (upper half) and the standard Yee solver (lower half). (a) Snapshot of the driving field $E_{y,A}$ (pulse A) and (b) $E_{z,B}$ (pulse B) inside the plasma. (c) Snapshot of the source term $\mathbf{E}\cdot\mathbf{B}$ in vacuum (before entering the plasma). (d) Axion field $\phi$ generated by the QDS solver. (e) Axion field $\phi$ generated by the Yee solver, showing degradation. (f) Temporal evolution of the conversion ratio (total axion energy over total laser energy). }
    \label{fig:two_color_mixing}
\end{figure}

Figure~\ref{fig:two_color_mixing} presents the benchmark results. Figures~\ref{fig:two_color_mixing}(a) and (b) show the two driving laser pulses co-propagating in the plasma. One difference between these two solvers appears in vacuum ($x<0$), shown in Fig.~\ref{fig:two_color_mixing}(c). Here, the physical source term $\mathbf{E}\cdot\mathbf{B}$ should vanish. The standard Yee solver (lower half), however, exhibits non-zero numerical noise due to the mismatch in numerical phase velocities $v_{p}(\omega_0) \neq v_{p}(2\omega_0)$. In contrast, the QDS solver (upper half) maintains the source term at zero.

The impact on axion generation is quantified in Fig.~\ref{fig:two_color_mixing}(f). In the vacuum region, the Yee solver produces an unphysical axion signal, whereas the QDS solver yields zero signal as physically required. Upon entering the plasma, the QDS solver maintains the correct phase matching conditions over long distances, resulting in the expected quadratic growth of the conversion ratio. The Yee solver, suffering from numerical dispersion, rapidly dephases from the resonance condition, causing the conversion efficiency to saturate prematurely. The resulting axion field distributions are compared in Fig.~\ref{fig:two_color_mixing}(d) for QDS and (e) for Yee, where the latter shows significantly weaker amplitude and unphysical structures. These results demonstrate that the cylindrical QDS solver is essential for accurately modeling sensitive phase-matched interactions in multi-frequency laser-plasma setups.

\section{Conclusion}\label{sec:conclusion}
We have generalized a quasi-cylindrical direction-splitting (QDS) dispersionless solver in the EPOCH framework, enabling electromagnetic and axion fields to propagate along the axial direction without numerical dispersion.
Benchmark simulations confirm the practical benefits of this formulation. The vacuum laser-propagation test shows that the QDS solver preserves the expected group velocity of a finite-waist pulse even at low axial resolution. In laser-wakefield runs, the cylindrical QDS solver reproduces the wakefield shape and amplitude of the standard Yee solver but preserves the correct laser group velocity at a low axial resolution, and also shortens the runtime by orders of magnitude compared to the fully 3D simulation. In axion-generation tests involving two-color laser mixing, the QDS solver correctly suppresses the unphysical noise in vacuum and captures the resonant quadratic growth of the axion field in the plasma, demonstrating its distinct advantage in modeling phase-sensitive axion--photon dynamics.
The algorithm provides a foundation for quasi-3D studies that require long-distance, dispersion-free laser propagation together with weak axion signals.

\vspace{2ex}
\noindent\textbf{Acknowledgments.}
This work was supported by the National Natural Science Foundation of China (Grants No. 12225505, 12090060, 12090061),
Fundamental and Interdisciplinary Disciplines Breakthrough Plan of the Ministry of Education of China (JYB2025XDXM204),
Office of Science and Technology and Shanghai Municipal Government (Grant No. 23JC1410200),
Shanghai Jiao Tong University 2030 Initiative, and partially supported by State Key Laboratory of Dark Matter Physics. We thank for the sponsorship from the Hongwen Foundation in Hong Kong and New Cornerstone Science Foundation in China. The computations in this paper were run on the $\pi$ 2.0 cluster supported by the Center for High Performance Computing at Shanghai Jiao Tong University.

\bibliographystyle{elsarticle-num}
\bibliography{dispersionless_solver_cylindrical}

@misc{an_enhanced_2025,
	title = {Enhanced axion-photon conversion via seeded photons instead of resonant cavities for short-pulse axion detection},
	language = {en},
	urldate = {2025-10-02},
	publisher = {arXiv},
	author = {An, Xiangyan and Chen, Min and Liu, Jianglai and Sheng, Zhengming and Zhang, Jie},
	month = sep,
	year = {2025},
	note = {arXiv:2509.21417 [hep-ph]},
}

@article{lehe_spectral_2016,
	title = {A spectral, quasi-cylindrical and dispersion-free particle-{In}-cell algorithm},
	volume = {203},
	issn = {00104655},
	language = {en},
	urldate = {2025-12-25},
	journal = {Comput. Phys. Commun.},
	author = {Lehe, Rémi and Kirchen, Manuel and Andriyash, Igor A. and Godfrey, Brendan B. and Vay, Jean-Luc},
	month = jun,
	year = {2016},
	pages = {66--82},
}

@article{van_bibber_proposed_1987,
	title = {Proposed experiment to produce and detect light pseudoscalars},
	volume = {59},
	copyright = {http://link.aps.org/licenses/aps-default-license},
	issn = {0031-9007},
	language = {en},
	number = {7},
	journal = {Phys. Rev. Lett.},
	author = {Van Bibber, K. and Dagdeviren, N. R. and Koonin, S. E. and Kerman, A. K. and Nelson, H. N.},
	month = aug,
	year = {1987},
	pages = {759--762},
}

@article{Sikivie1983,
  author = {Pierre Sikivie},
  title = {Experimental Tests of the "Invisible" Axion},
  journal = {Phys. Rev. Lett.},
  volume = {51},
  number = {16},
  pages = {1415--1417},
  year = {1983}
}

@article{Fonseca2008,
  author = {R. A. Fonseca and S. F. Martins and L. O. Silva and J. W. Tonge and F. S. Tsung and W. B. Mori},
  title = {One-to-one direct modeling of experiments and astrophysical scenarios: pushing the envelope on kinetic plasma simulations},
  journal = {Plasma Phys. Control. Fusion},
  volume = {50},
  number = {12},
  pages = {124034},
  year = {2008}
}

@article{Pukhov2016,
  author = {Alexander Pukhov},
  title = {Particle-in-cell codes for plasma-based particle acceleration},
  journal = {CERN Yellow Rep.},
  volume = {1},
  pages = {181--199},
  year = {2016}
}

@article{Vay2013,
  author = {Jean-Luc Vay and Irving Haber and Brendan B. Godfrey},
  title = {A domain decomposition method for pseudo-spectral electromagnetic simulations of plasmas},
  journal = {J. Comput. Phys.},
  volume = {243},
  pages = {260--268},
  year = {2013}
}

@article{Derouillat2018,
  author = {Julien Derouillat and Antoine Beck and Fr{\'e}d{\'e}ric P{\'e}rez and Thierry Vinci and Matthieu Chiaramello and Alexandre Grassi and Marica L{\'e} and Guillaume Bouchard and Ilya Plotnikov and Nicolas Aunai and Jordan Dargent and Christine Riconda and Matthieu Grech},
  title = {SMILEI: a collaborative, open-source, multi-purpose particle-in-cell code for plasma simulation},
  journal = {Comput. Phys. Commun.},
  volume = {222},
  pages = {351--373},
  year = {2018}
}

@article{Burau2010,
  author = {Heiko Burau and Renee Widera and Wolfgang H{\"o}nig and Guido Juckeland and Alexander Debus and Thomas Kluge and Michael Bussmann},
  title = {PIConGPU: a fully relativistic particle-in-cell code for a GPU cluster},
  journal = {IEEE Trans. Plasma Sci.},
  volume = {38},
  number = {10},
  pages = {2831--2839},
  year = {2010}
}

@article{an_modeling_2024,
	title = {Modeling of axion and electromagnetic fields interaction in particle-in-cell simulations},
	volume = {9},
	issn = {2468-2047, 2468-080X},
	language = {en},
	number = {6},
	journal = {Matter and Radiation at Extremes},
	author = {An, Xiangyan and Chen, Min and Liu, Jianglai and Sheng, Zhengming and Zhang, Jie},
	month = nov,
	year = {2024},
	pages = {067204},
}

@article{Pukhov2020,
  author    = {Alexander Pukhov},
  title     = {X-dispersionless Maxwell solver for plasma-based particle acceleration},
  journal   = {J. Comput. Phys.},
  volume    = {418},
  pages     = {109622},
  year      = {2020}
}

@article{Yee1966,
  author    = {Kane Yee},
  title     = {Numerical solution of initial boundary value problems involving Maxwell's equations in isotropic media},
  journal   = {IEEE Trans. Antennas Propag.},
  volume    = {14},
  number    = {3},
  pages     = {302--307},
  year      = {1966}
}

@article{Godfrey2013,
  author    = {Brendan B. Godfrey and Jean-Luc Vay},
  title     = {Numerical stability of relativistic beam multidimensional PIC simulations employing the Esirkepov algorithm},
  journal   = {J. Comput. Phys.},
  volume    = {248},
  pages     = {33--46},
  year      = {2013}
}

@article{Lehe2016,
  author    = {R. Lehe and M. Kirchen and B.~B. Godfrey and A.~R. Maier and J.-L. Vay},
  title     = {Elimination of numerical Cherenkov instability in flowing-plasma {PIC} simulations by using Galilean coordinates},
  journal   = {Phys. Rev. E},
  volume    = {94},
  pages     = {053305},
  year      = {2016}
}

@article{Lehe2013,
  author    = {R. Lehe and A. Lifschitz and C. Thaury and V. Malka and X. Davoine},
  title     = {Numerical growth of emittance in simulations of laser-wakefield acceleration},
  journal   = {Phys. Rev. ST Accel. Beams},
  volume    = {16},
  pages     = {021301},
  year      = {2013},
  doi       = {10.1103/PhysRevSTAB.16.021301}
}

@article{Giovannini2015,
  author    = {Daniel Giovannini and Jacquiline Romero and V{\'a}clav Poto{\v{c}}ek and Gergely Ferenczi and Fiona Speirits and Stephen M. Barnett and Daniele Faccio and Miles J. Padgett},
  title     = {Spatially structured photons that travel in free space slower than the speed of light},
  journal   = {Science},
  volume    = {347},
  number    = {6224},
  pages     = {857--860},
  year      = {2015},
  doi       = {10.1126/science.aaa3035}
}

@article{Nuter2014,
  author    = {R. Nuter and M. Grech and P. Gonzalez De Alaiza Martinez and G. Bonnaud and E. d'Humi{\`e}res},
  title     = {Maxwell solvers for the simulations of the laser-matter interaction},
  journal   = {Eur. Phys. J. D},
  volume    = {68},
  pages     = {177},
  year      = {2014}
}

@article{Arber2015,
  author    = {T.~D. Arber and K. Bennett and C.~S. Brady and A. Lawrence-Douglas and M.~G. Ramsay and N.~J. Sircombe and P. Gillies and R.~G. Evans and H. Schmitz and A.~R. Bell and C.~P. Ridgers},
  title     = {Contemporary particle-in-cell approach to laser-plasma modelling},
  journal   = {Plasma Phys. Controlled Fusion},
  volume    = {57},
  number    = {11},
  pages     = {113001},
  year      = {2015}
}

@article{Lifschitz2009,
  author    = {A.~F. Lifschitz and X. Davoine and E. Lefebvre and J. Faure and C. Rechatin and V. Malka},
  title     = {Particle-in-cell modeling of laser--plasma interaction using Fourier decomposition},
  journal   = {J. Comput. Phys.},
  volume    = {228},
  pages     = {1803--1814},
  year      = {2009}
}

@inproceedings{Boris1970,
  author    = {J.~P. Boris},
  title     = {Relativistic plasma simulation -- optimization of a hybrid code},
  booktitle = {Proc. Fourth Conf. on Numerical Simulation of Plasmas},
  pages     = {3--67},
  address   = {Naval Research Lab, Washington, D.C.},
  year      = {1970}
}

@article{Esirkepov2001,
  author    = {Y.~A. Esirkepov},
  title     = {Exact charge conservation scheme for Particle-in-Cell simulation with an arbitrary form-factor},
  journal   = {Comput. Phys. Commun.},
  volume    = {135},
  pages     = {144--153},
  year      = {2001}
}

@article{Vay2011,
  author    = {Jean-Luc Vay},
  title     = {Effects of hyperbolic rotation in Minkowski space on the modeling of plasma accelerators in a Lorentz boosted frame},
  journal   = {Phys. Plasmas},
  volume    = {18},
  number    = {3},
  pages     = {030701},
  year      = {2011}
}

@misc{cylindrical_epoch_github,
  howpublished          = {https://github.com/epochpic/cylindrical\_epoch}
}

@article{Peccei1977a,
  author  = {R. D. Peccei and H. R. Quinn},
  title   = {CP Conservation in the Presence of Pseudoparticles},
  journal = {Phys. Rev. Lett.},
  volume  = {38},
  pages   = {1440--1443},
  year    = {1977}
}

@article{Peccei1977b,
  author  = {R. D. Peccei and H. R. Quinn},
  title   = {Constraints Imposed by CP Conservation in the Presence of Pseudoparticles},
  journal = {Phys. Rev. D},
  volume  = {16},
  pages   = {1791--1797},
  year    = {1977}
}

@article{Weinberg1978,
  author  = {Steven Weinberg},
  title   = {A New Light Boson?},
  journal = {Phys. Rev. Lett.},
  volume  = {40},
  pages   = {223--226},
  year    = {1978}
}

@article{Wilczek1978,
  author  = {Frank Wilczek},
  title   = {Problem of Strong {P} and {T} Invariance in the Presence of Instantons},
  journal = {Phys. Rev. Lett.},
  volume  = {40},
  pages   = {279--282},
  year    = {1978}
}

@article{Preskill1983,
  author  = {J. Preskill and M. B. Wise and F. Wilczek},
  title   = {Cosmology of the Invisible Axion},
  journal = {Phys. Lett. B},
  volume  = {120},
  pages   = {127--132},
  year    = {1983}
}

@article{Abbott1983,
  author  = {L. F. Abbott and P. Sikivie},
  title   = {A Cosmological Bound on the Invisible Axion},
  journal = {Phys. Lett. B},
  volume  = {120},
  pages   = {133--136},
  year    = {1983}
}

@article{Dine1983,
  author  = {M. Dine and W. Fischler},
  title   = {The Not So Harmless Axion},
  journal = {Phys. Lett. B},
  volume  = {120},
  pages   = {137--141},
  year    = {1983}
}

@article{Tercas2018,
  author  = {H. Tercas and J. Rodrigues and J. T. Mendonca},
  title   = {Axion-Plasmon Polaritons in Strongly Magnetized Plasmas},
  journal = {Phys. Rev. Lett.},
  volume  = {120},
  pages   = {181803},
  year    = {2018}
}

@article{Burton2018,
  author  = {D. A. Burton and A. Noble},
  title   = {Plasma-Based Wakefield Accelerators as Sources of Axion-Like Particles},
  journal = {New J. Phys.},
  volume  = {20},
  pages   = {033022},
  year    = {2018}
}

@article{Mendonca2020a,
  author  = {J. T. Mendonca and H. Tercas and J. D. Rodrigues},
  title   = {Axion Excitation by Intense Laser Fields in a Plasma},
  journal = {Phys. Scr.},
  volume  = {95},
  pages   = {045601},
  year    = {2020}
}

@article{Mendonca2020b,
  author  = {J. Mendonca and J. Rodrigues and H. Tercas},
  title   = {Axion Production in Unstable Magnetized Plasmas},
  journal = {Phys. Rev. D},
  volume  = {101},
  pages   = {051701},
  year    = {2020}
}

@article{Huang2022,
  author  = {S. Huang and B. Shen and Z. Bu and X. Zhang and L. Ji and S. Zhai},
  title   = {Axion-Like Particle Generation in Laser-Plasma Interaction},
  journal = {Phys. Scr.},
  volume  = {97},
  pages   = {105303},
  year    = {2022}
}

@misc{an_situ_2025,
	title = {In situ axion generation and detection in laser-plasma wakefield interaction},
	language = {en},
	publisher = {arXiv},
	author = {An, Xiangyan and Chen, Min and Liu, Jianglai and Bai, Zhan and Ji, Liangliang and Sheng, Zhengming and Zhang, Jie},
	year = {2025},
	note = {arXiv:2504.12500 [physics]},
}

\end{document}